\documentclass{PoS}

\usepackage{graphicx}
\newcommand{\la}{\langle}
\newcommand{\ra}{\rangle}

\newcommand{\beq}{\begin{eqnarray}}
\newcommand{\eeq}{\end{eqnarray}}
\newcommand{\sbeq}{\begin{subeqnarray}}
\newcommand{\seeq}{\end{subeqnarray}}

\newcommand{\bl}{\biggl}
\newcommand{\br}{\biggr}
\newcommand{\bfk}{{\bf k}}

\newcommand{\bfv}{{\bf v}}

\title{Toward Identifying the QCD Critical Point: 
attenuation of the sound mode around the critical point }

\ShortTitle{Identifying the QCD Critical Point}

\author{\speaker{Teiji Kunihiro} and Yuki Minami\\
        Department of Physics, Kyoto University, Kyoto 606-8502, Japan\\
 E-mail: \email{kunihiro@ruby.scphys.kyoto-u.ac.jp}, \email{y-minami@ruby.scphys.kyoto-u.ac.jp}}

\abstract{
Motivated by the fact that QCD critical point (CP) belongs to
the same universality class as the liquid-gas transition,
the dynamical density fluctuations around the CP is analyzed using relativistic
fluid dynamics for a viscous fluid.
It is shown that   relativistic effects on the dynamical structure factor 
of the density fluctuation appears only in the width of the  sound and thermal modes
through a modification of the transport/kinetic coefficients.
It is found that 
 the mechanical density fluctuation
which is enhanced by the relativistic effects are attenuated,
whereas the entropy fluctuation in turn 
becomes more prominent around the QCD CP.
This attenuation of the density fluctuation around the QCD CP may imply
 that if the suppression or even total disappearance
 of  Mach cone is observed as the incident energy of the relativistic heavy-ion collisions
is lowered, it can be a signal of the existence of  the QCD critical point.
}

\FullConference{5th International Workshop on Critical Point and Onset of Deconfinement - CPOD 2009,\\
		 June 08 - 12 2009\\
		 Brookhaven National Laboratory, Long Island, New York, USA}

\begin{document}

\section{Introduction}

An interesting  feature of the QCD phase diagram
 is the existence of a critical point\cite{qcdcp,Ejiri:2007ga};
see Refs. \cite{Zhang:2008wx,Hatsuda:2006ps,Fukushima:2008is} 
for possible variants and alternatives.
At the QCD critical point (CP), the first order phase transition terminates and turns to 
a second order phase transition. 
It is now known that 
the QCD CP belongs to the universality class H in 
the classification scheme by Hohengerg-Halperin\cite{Hohenberg:1977ym},
i.e., the same as that of 
the liquid-gas phase transition point\cite{fujii,son}.
This implies that the density fluctuating mode 
 and, generically, fluid dynamic
modes coupled to conserved quantities
are the  softening modes at the CP.

The $\sigma$ mode that is coupled to the fluctuation
of the chiral order parameter $\la \bar{q}q)\ra$
becomes the 
soft mode of the chiral transition at finite temperature $T$ but vanishing chemical
potential $\mu=0$.
The QCD CP exists, however,
 at finite chemical potential $\mu$ and for finite current quark masses.
In such a case,
as is first shown in \cite{Kunihiro:1991qu},
the scalar-vector cross term $\la :(\bar{q}\gamma^0q)(\bar{q}q):\ra$  does not vanish,
and, hence, the scalar mode  is coupled with
the density-density correlator $\la :(\bar{q}\gamma^0q)^2:\ra$;
this is because charge conjugation symmetry is violated with finite $\mu$,
and the left- and right-handed quarks are coupled owing to the breaking
of chiral symmetry.\footnote{
It is also notable that the density-density correlator is a
part of the polarization tensor $\la :(\bar{q}\gamma^{\mu}q)(\bar{q}\gamma^{\nu}q):\ra$,
which suggests that the vector coupling $g_{_V}(\bar{q}\gamma^{\mu}q)^2$ plays
an important role for the static and dynamic properties of the QCD CP.\cite{Kunihiro:1991qu}
}
Then, the would-be soft mode, the $\sigma$ mode, remains massive (a fast mode) and 
becomes a slaving mode of the genuine soft mode given 
by the fluid dynamical modes coupled to the density fluctuation\cite{fujii,son}.
Here we note that the diverging behavior of the density fluctuation
around the QCD CP was first shown in \cite{Kunihiro:2000ap}; see also \cite{Hatta:2002sj}.

Recently, it has been argued \cite{Moore} on the basis of the analysis 
of the liquid-gas phase transition by Onuki\cite{onuki:1997} that 
the bulk viscosity may show a singular behavior around the QCD CP.

We explore how the singularities of the thermodynamic
quantities as well as the transport coefficients affect the dynamical density
 fluctuations around the QCD CP thereby we can have a hint how to 
identify the signal of the existence of QCD CP by experiment\cite{Minami:2009hn}.
Our investigation is  based on an explicit use
of the {\em relativistic} fluid dynamic
equations for a viscous fluid.
We also demonstrate that even the so called first-order 
relativistic fluid dynamic equations have generically no problem
to describe fluid dynamical phenomena with long wave lengths contrary to
a possible naive suspect.

We find that
 the mechanical density fluctuation
which is enhanced by the relativistic effects are attenuated,
but the entropy fluctuation in turn 
becomes more prominent around the QCD CP,
which is, to our surprise, precisely the same in effect as
the critical behavior  in  the nonrelativistic case.\cite{stanley}
On the basis of these findings,
we speculate that if  a Mach cone, which originates from the dynamical
density fluctuations stimulated by a fast particle and seems to have been
observed in RHIC experiment\cite{:2008nd},
 disappears or is strongly suppressed 
as the incident energy of relativistic heavy-ion collisions is lowered,
it can be a signal of the existence of the QCD CP provided that 
the incident energy is still high enough to  make jets\cite{Torrieri:2009mv}.

\section{Dissipative Relativistic Fluid Dynamics}

Relativistic fluid dynamic equations consist of
the balance equations for energy-momentum and particle number,   
%\begin{eqnarray}
$\partial_\mu T^{\mu \nu}=0$ and 
$\partial_\mu N^\mu =0$
%\end{eqnarray}
where $T^{\mu \nu}$ is the energy-momentum tensor
and  $N^\mu$  the particle current:
\begin{eqnarray}
T^{\mu \nu}=(\epsilon+P)u^{\mu}u^{\nu}-Pg^{\mu\nu}+\tau^{\mu\nu},  \quad 
N^\mu  = n u^\mu+\nu^\mu. 
\end{eqnarray}
Here $\epsilon$ is the energy density, $P$  the pressure, $u^\mu$ the flow velocity, 
and $n$ the particle density; the dissipative part of the energy-momentum
tensor and the particle current are denoted by $\tau^{\mu \nu}$ and $\nu^\mu$,
respectively.

The so called first-order equations such as Landau\cite{landau} and Eckart\cite{eckart}
ones are parabolic and formally violates the causality,
and, hence, are called acausal.
Moreover, the Eckart equation which is defined for the particle frame
where the particle current does not have a dissipative part 
($\nu^{\mu}=0$) shows a pathological
 property that the fluctuations around the thermal equilibrium is
unstable\cite{hiscock}.
% while the Landau equation defined for the energy frame does
%not show such a pathological behavior. 
The causality problem is circumvented in the Israel-Stewart equation\cite{is},
which is a second-order equation with relaxation times incorporated.

One should, however, notice
that formally acausal fluid dynamic equations should be
valid in describing
 fluid dynamical phenomena with longer wavelengths than the mean free path.
In fact, we will see that the results for fluid dynamical modes
with long wave lengths are qualitatively the same 
irrespective whether the second-order or first-order equations are
used.

As for the instability seen in the Eckart equation, a new first-order
equation in the particle frame constructed by Tsumura, Kunihiro and Ohnishi (TKO)
\cite{tko} has no such a pathological behavior\cite{tk}. 
We employ Landau\cite{landau}, Eckart\cite{eckart} and Israel-Stewart(I-S)\cite{is}
 equation as typical equations,
and TKO equation in particle frame.

\section{Dynamic structure factor of density fluctuations}

In this section,
we first present 
a procedure for deriving the dynamical structure factor (spectral function)
of the density fluctuation
for the Landau equation. 
Then we just give the results for Eckart, Tsumura-Kunihiro-Ohnishi(TKO)
equation and the Israel-Stewart equation in particle frame.

\subsection{Landau equation (energy frame)}

The dissipative terms in the Landau equation are given by
\begin{eqnarray}
\tau^{\mu \nu}=\eta [\partial^{\mu}_{\perp}u^{\nu}
               +\partial^{\nu}_{\perp}u^{\mu}
               -\frac{2}{3}\Delta^{\mu\nu}(\partial_{\perp}{\cdot}u)]
               +\zeta\Delta^{\mu\nu}(\partial_{\perp}{\cdot}u),\,  \quad 
\nu^{\mu} = \kappa \bl( \frac{n T}{w} \br)^2 \partial_{\perp}^\mu \bl(\frac{\mu}{T} \br),
\end{eqnarray} 
where $\eta$ is the shear viscosity, $\zeta$  the bulk viscosity, 
$\kappa$  the thermal conductivity and $w=\epsilon+P$ the enthalpy density.
 $\Delta^{\mu \nu}=g^{\mu \nu}-u^\mu u^\nu$ is the projection operator 
on the space-like vector, and 
$\partial_{\perp}^{\mu}=\Delta^{\mu \nu}\partial_{\nu}$ 
the space-like derivative (gradient operator). 

We calculate the dynamical structure factor,
or the spectral function, of the  density fluctuation 
around the thermal equilibrium state, as was done for
 the non-relativistic case\cite{mountain,reichl};
 we refer to  the comprehensive text book 
Ref.~\cite{reichl}.

Let us  express the dynamical quantities
as follows,
 $n(x)=n_0+\delta n(x)$, 
$\epsilon(x)=\epsilon_0 + \delta \epsilon (x)$, $P(x)=P_0+\delta P(x)$, 
$\mu(x)=\mu_0 + \delta \mu (x)$, and $u^\mu (x)=u^\mu_0 + \delta u^\mu (x)$,
where the variables with the subscript $_0$ denote
those in the equilibrium state in at rest;
$u_0^\mu=(1,\bf{0})$. Then,
 $\delta u^\mu (x)= (0, \delta \bfv (x))$
with $\delta \bfv (x)$ yet to be determined together with other quantities
like $\delta n$ etc. 
Now we have five equations for seven unknown quantities, $\delta n$, $\delta T$,
$\delta P$, $\delta s$, and $\delta \bfv $.
Choosing $\delta n$ and  $\delta T$ as independent variables,  we have
$\delta P(x)=\frac{w_{0} c_{s}^{2}}{n_{0} \gamma}\delta n(x)
            + \frac{w_{0}c_{s}^{2}\alpha_{P}}{\gamma}\delta T(x)$\,
and 
$\delta s(x)=-\frac{w_{0}c_{s}^{2}\alpha_{P}}{n_{0}^{2}\gamma}\delta n(x)
            + \frac{\tilde c_{n}}{T_{0}} \delta T(x)$.
Here $\tilde{c}_n=T_0(\partial s / \partial T)_n$ and $\tilde{c}_P=T_0(\partial s /
 \partial T)_P$
are the specific heats at constant density and pressure, respectively, 
$c_s=(\partial P/\partial \epsilon)_s^{1/2}$ the sound velocity, 
$\alpha_P=$\, $-(1/n_0)(\partial n / \partial T)_P$  the thermal expansivity 
at constant pressure, and $\gamma=\tilde{c}_P/\tilde{c}_n$  the ratio of the specific
 heats.   
The equation describing the density fluctuations is finally found to be
\begin{eqnarray}
\{ \frac{\partial}{\partial t}-\kappa\frac{T_0 c_s^2}{w_0 \gamma}\nabla^2 \} \delta n
 +n_0\nabla\cdot\delta\bfv 
 +\kappa\frac{n_0}{w_0}(1-\frac{c_s^2 \alpha_P T_0}{\gamma})\nabla^2 \delta T=0, \\
w_{0}\frac{\partial \delta\bfv }{\partial t}-\eta\nabla^{2}\delta\bfv 
 -(\zeta+\frac{1}{3}\eta)\nabla(\nabla\cdot\delta\bfv )
 +\frac{w_0 c_s^2}{n_0\gamma}\nabla \delta n
 +\frac{w_0 c^2_s \alpha_P}{\gamma}\nabla \delta T=0,                        \\
\{ -\frac{w_0 c_s^2 \alpha_P }{n_0 \gamma }\frac{\partial }{\partial t}
  +\kappa \frac{c_s^2}{n_0\gamma } \nabla^2  \} \delta n
 +\{ \frac{n_0\tilde{c}_n}{T_0} \frac{\partial }{\partial t}
  +\kappa(\frac{c_s^2\alpha_P }{\gamma }-\frac{1}{T_0})\nabla^2 \} \delta T=0. 
\end{eqnarray}

Now we are interested in the dynamical structure factor of the density fluctuation
 as given by 
\beq
S_{n n}(\bfk ,\omega) \equiv \la \delta \tilde{n}(\bfk ,\omega) \delta n (\bfk ,t=0)\ra,
\eeq
where $ \delta \tilde{n}(\bfk, \omega) $ 
 is the Fourier transform of the density fluctuation, 
and $\la\,\,\ra$ denotes the thermal average in the equilibrium. 

Note that $\delta T$ and $\delta n$ are statistically independent in fluid systems
which is established in Einstein fluctuation theory\cite{reichl};
$\la \delta T(\bfk ,0)\delta n(\bfk ,0)\ra =0$, and similarly,
$\la \delta v_{\parallel}(\bfk ,0)\delta n(\bfk ,0)\ra =0 $,
where $\delta v_{\parallel}(\bfk ,0)$ denotes the longitudinal
component of the velocity field.
Thus we have the dynamical structure factor
\begin{eqnarray}
\frac{S_{n n}(\bfk ,\omega )}{\la (\delta n(\bfk ,t=0))^2\ra}    =  (1-\frac{1}{\gamma})
   \frac{2\Gamma_{\rm R} k^{2}}{\omega^{2}+\Gamma_{\rm R}^{2}k^{4}}
   +\frac{1}{\gamma}
   \{\frac{\Gamma_{\rm B} k^{2}}{(\omega -c_{s}k)^{2}+\Gamma_{\rm B}^{2}k^{4}}
   +\frac{\Gamma_{\rm B} k^{2}}{(\omega +c_{s}k)^{2}+\Gamma_{\rm B}^{2}k^{4}}\} .
   \label{eq:landau}
\end{eqnarray}
Here,
$\Gamma_{\rm R}=\kappa/(n_0\tilde{c}_P)$ and
$\Gamma_{\rm B}=\frac{1}{2}[\chi(\gamma-1)+\nu_l ]$
$+\frac{1}{2}c_s^2 T_0 (\frac{\kappa}{w_0} -2\chi \alpha_P )$
with $\nu_l=(\zeta+4\eta/3)/w_0$ being the ({\em relativistic})
longitudinal kinetic viscosity; notice $w_0$ in the denominator.
 We also remark that the width
$\Gamma_{\rm R}$ is identified with the thermal diffusivity $\chi$.
 
We see that the spectral function has three peaks at frequencies $\omega=0$ and 
$\omega = \pm c_s k$:
The peak at $\omega=0$ corresponds to thermally induced density fluctuations,
which is called  Rayleigh peak;  while
the two-side peaks at $\omega=\pm c_s k$ correspond to mechanically induced density 
fluctuation, i.e. sound waves.
These two peaks are called Brillioun peaks.  

We find that  relativistic effects 
manifest themselves only in the width of the Brillouin  peaks $\Gamma_{\rm B}$, whereas
the width of the Rayleigh peak 
is the same as the non-relativistic case\cite{mountain,reichl}.
The relativistic effects in $\Gamma_{\rm B}$ appear in two ways, as seen
by the expression,
\beq
\Gamma_{\rm B}=\Gamma_{\rm B}^{\rm MR}+\delta \Gamma_{\rm B}^{\rm La}
\eeq
where
$\Gamma_{\rm B}^{\rm MR} \equiv \frac{1}{2}[\chi(\gamma-1)+\nu_l ]$
and
$\delta \Gamma_{\rm B}^{\rm La} \equiv  \frac{1}{2}c_s^2 T_0 (\kappa / w_0 -2\chi \alpha_P )$.
Firstly, the first term has a nonrelativistic counter part
only with a replacement of the enthalpy
density $w_0$ with the mass density $\rho_0$ in $\nu_l$.
We call this modification the minimal relativistic (MR) effect.
On the other hand, the second term $\delta \Gamma_{B}^{\rm La}$
is  a genuine relativistic effect 
that is absent in the non-relativistic case; this comes from 
the mass-energy equivalence inherent in the relativistic theory.

\begin{figure}[!t]
 \begin{tabular}{cc}
\begin{minipage}{0.5\hsize}
  \begin{center}
   \includegraphics[width=50mm,angle=-90]{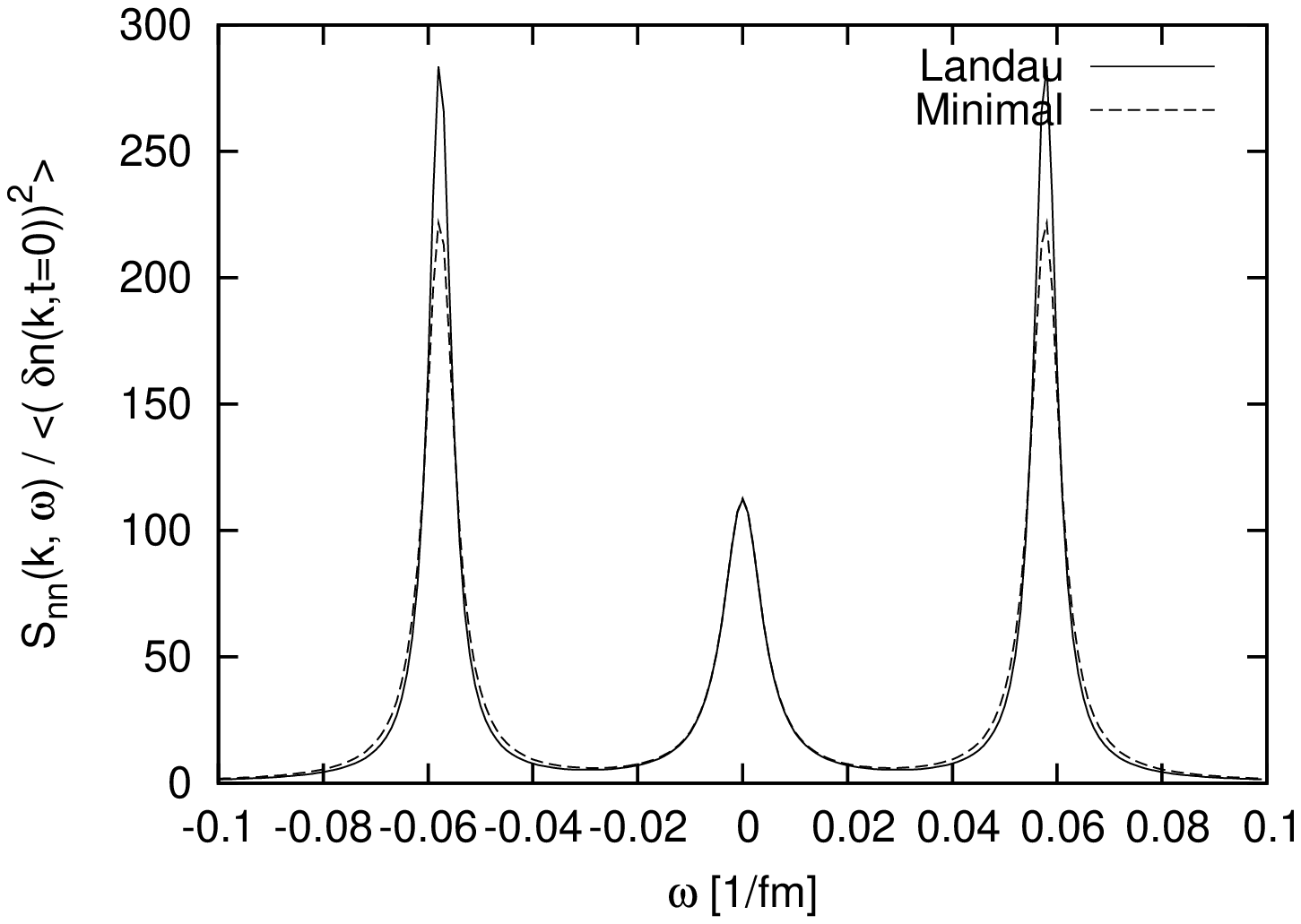}
\end{center}
\end{minipage}
\begin{minipage}{0.5\hsize}
 \begin{center}
   \includegraphics[width=50mm,angle=-90]{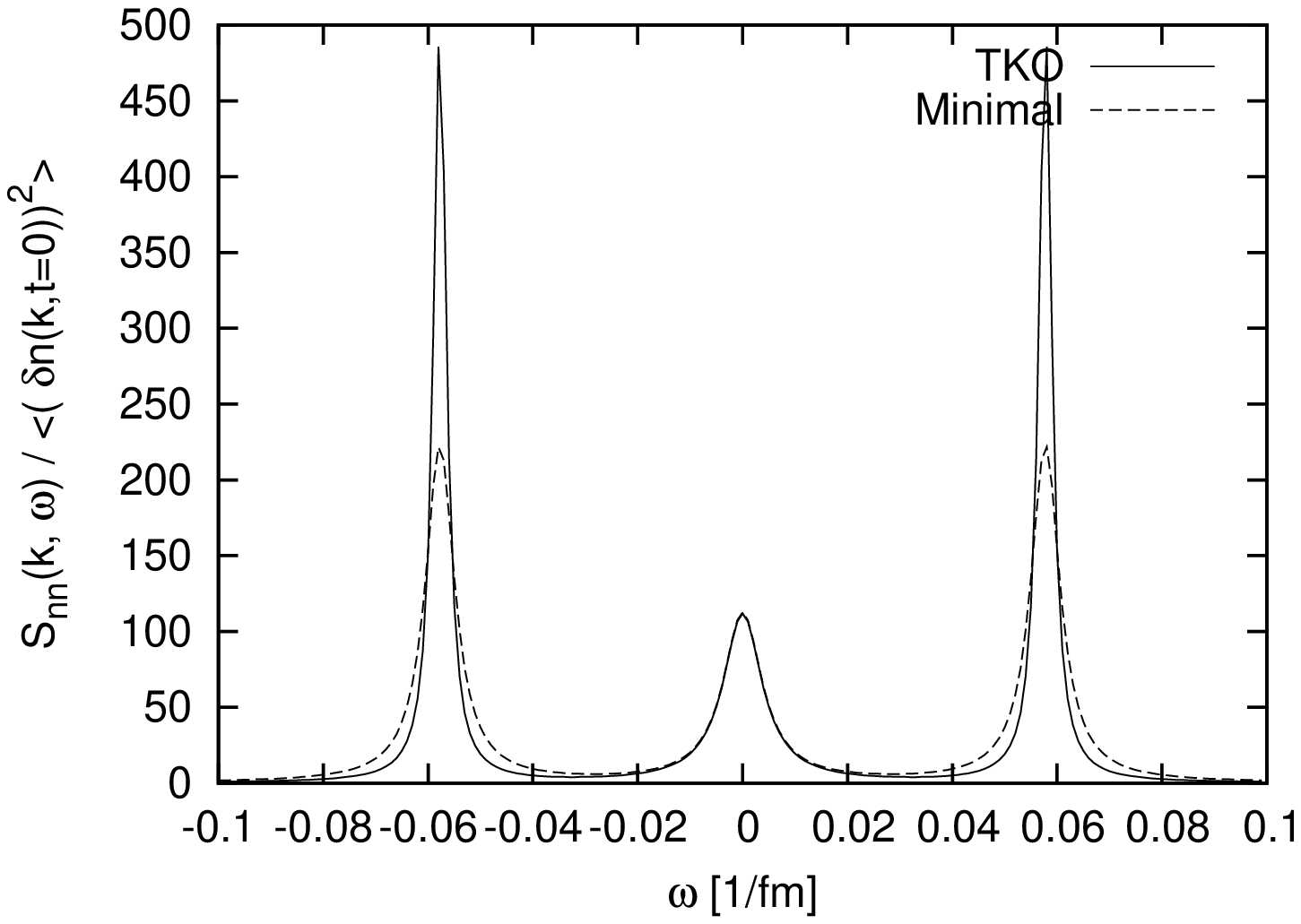}
\end{center}
\end{minipage}
\end{tabular}
\begin{center}
 %\begin{minipage}{1.0\hsize}
    \caption{Left panel: The solid line is
 the spectral function for the density
  fluctuations from
 Landau/Israel-Stewart equation with
  $k=0.1$[1/fm],\, $\mu_0=200$[MeV],\, 
           $T_0=200$[MeV],\, $\eta/(n_0 s_0)=\zeta/(n_0 s_0)=0.3$ and
           $\kappa T_0/(n_0 s_0)=0.6$; the dashed line
in the  minimal relativistic case. Right panel: the result from TKO
  equation with the same parameters. Notice that the vertical scale is
different from that of the left panel.
     }
  \label{fig:landau}
 %\end{minipage}
\end{center}
\end{figure}

Figure \ref{fig:landau} shows the spectral function Eq.(\ref{eq:landau}) and 
 the minimal relativistic case with $\delta \Gamma _B=0$ for  the parameter set\, 
$k=0.1$[1/fm],\, $\mu_0=200$[MeV],\, 
$T_0=200$[MeV],\, $\eta/(n_{0}s_0)=\zeta/(n_0 s_0)=0.3$ and $\kappa T_0/(n_0 s_0)=0.6$.
Note that $n_0 s_0$ represents the entropy density in the equilibrium state 
because $s_0$ is the entropy per particle number.   
As is expected,  Fig.\ref{fig:landau} shows that
the Brillouin peaks owing to the sound mode
 is enhanced by the relativistic effects,
while the Rayleigh peak owing to the thermal mode 
is the same as in the non-relativistic case. 

\subsection{Stable first-order fluid dynamic equation in the particle frame}

%A new fluid dynamic equation was derived from the relativistic 
%Boltzmann equation by 
% Tsumura, Kunihiro and Ohnishi (TKO)\cite{tko}, which
%is found to be stable in the sense that any fluctuations
%around the thermal equilibrium state relax down to recover the 
%equilibrium\cite{tk}.
%So let us take  TKO equation in the particle frame,
% and examine whether the spectral function of the density fluctuations
%or the dynamical structure factor shows any frame dependence.

In the case of TKO equation\cite{tko} in the particle frame ($\nu^{\mu}=0$),
  the dissipative terms are given by
\begin{eqnarray}
\label{eq:tkoeq}
\tau^{\mu\nu}=\eta [\,\partial^{\mu}_{\perp}u^{\nu}+
   \partial^{\nu}_{\perp}u^{\mu}-\frac{2}{3}\Delta^{\mu\nu}(\partial_{\perp}{\cdot}u)\,]
   -\zeta^{'}(3 u^{\mu}u^{\nu}-\Delta^{\mu\nu})(\partial_{\perp}{\cdot}u) 
%\nonumber \\
   +\kappa(u^{\mu}\partial^{\nu}_{\perp}T+u^{\nu}\partial _{\perp}^{\mu}),
%\nu^{\mu}=&0&,
\end{eqnarray}
where $ \zeta^{'} =\zeta/(3\gamma-4)^2$.
From the same procedure as taken for the Landau equation, 
we obtain the dynamical structure factor with the same form as 
(\ref{eq:landau}) but with a different form of the width of the 
Brillouin peaks\cite{Minami:2009hn};
\begin{equation}
\Gamma_{\rm B}=\frac{1}{2}[\chi(\gamma-1)+\nu^{\rm TKO}_{l}
  -\frac{\alpha_{P}c_{s}^{2}}{n_{0}\tilde{c}_{P}}
(\kappa T_{0}+3\zeta^{'})] 
\equiv \Gamma_{\rm B}^{\rm TKO}.
\end{equation}
We emphasize  that since the fluid dynamic  fluctuations around the equilibrium 
state is relaxing in TKO equation, we have obtained the dynamical
structure factor
 without any obstruction, in contrast to the case of 
Eckart equation\cite{eckart},
 for which we have found\cite{Minami:2009hn} that the unstable behavior of
the density fluctuation around the thermal equilibrium prevents us
from obtaining the dynamical structure factor 
in a sensible way.
 
We see that the genuine relativistic effect 
in the width of the Brillouin peaks reads
$\delta \Gamma_{\rm B}^{\rm TKO} \equiv 
-\frac{\alpha_{P}c_{s}^{2}}{2n_{0}\tilde{c}_{P}}(\kappa T_{0}+3\zeta^{'})$,
which is definitely negative, implying that  the relativistic effect acts to 
enhance and sharpen the spectral function of the density fluctuation
in comparison with the minimal relativistic case than in 
the energy frame; see the right panel in Fig.\ref{fig:landau}; notice that
the scales of the right and left panel are different.
%See, however, the next subsection 
%for the Israel-Stewart equation
%in the particle frame.

\subsection{Israel-Stewart equation}

The derivation of the dynamical structure factor 
for the Israel-Stewart (I-S) equation\cite{is} can be performed
much the same way as for the Landau equation with some 
additional complications
due to the presence of the relaxation times.
However, it is found that 
 if the relaxation time $\beta_1$ of the heat current is
so small as to satisfy the condition
$\beta_1 < \frac{1}{w_0}$,
the I-S equation in particle frame takes over the pathological 
behavior of Eckart one even though the relaxation time is finite;
i.e., the density fluctuation will not relax down.

If we  assume that the relaxation time is sufficiently large
to satisfy the inequality $\beta_1 > \frac{1}{w_0}$,
then we can have the spectral function,
\begin{eqnarray}
S_{n n}(\bfk ,\omega )/{\la (\delta n(\bfk ,t=0))^2\ra }=
  &&(1-\frac{1}{\gamma})\frac{2\chi k^{2}}{\omega^{2}+\chi^{2}k^{4}} 
   +\frac{1}{\gamma}[\frac{\Gamma_{\rm B} k^{2}}{(\omega -c_{s}k)^{2}+\Gamma_{\rm B}^{2}k^{4}} \nonumber \\
     &&+\frac{\Gamma_{\rm B} k^{2}}{(\omega +c_{s}k)^{2}+\Gamma_{\rm B}^{2}k^{4}}]      
     +O(k^2) \times [\frac{2  /\beta_0\zeta}{\omega^{2}+1/(\beta_0\zeta )^2}  \nonumber \\  
       &&+\frac{1  /\beta_2\eta}{\omega^{2}+1/(2\beta_2\eta)^2}
       +\frac{2 w_0 /[(\beta_1 w_0-1)\kappa T_0]}
       {\omega^{2}+w_0^2 /[(\beta_1 w_0-1)\kappa T_0]^2} ].
 \label{eq:is}
\end{eqnarray}
Here,  $\beta_0 $ and $\beta_2 $ denote the relaxation times for
the bulk viscosity and  the shear viscosity, respectively.
Apparently the spectral function has six peaks including the conventional
three peaks,
but the new  three Lorentzian functions should vanish in the long wavelength limit 
$k \rightarrow 0$,
because the strength of these is in the second order of $k$.
Therefore Israel-Stewart equation gives completely 
 the same result for the dynamical structure factor
for the density fluctuations
in the long wavelength limit  as that Landau equation gives.
That is, the relaxation times do not affect the result in the 
fluid dynamical regime.

\section{The behavior around the QCD critical point}

We are now in a position to
 analyze the behavior of the spectral function of the density fluctuations
around the QCD critical point,
 on the basis of the dynamic as well as static
scaling laws for the liquid-gas transition \cite{stanley,onuki:1997}.

The specific heat at constant density $\tilde{c}_n$ and the 
the isothermal compressibility 
$K_T=\frac{1}{n_0}$$\big(\frac{\partial n}{\partial P}\big)_T$
 diverge at the critical point as
$\tilde{c}_n = c_0 t^{-\tilde{\alpha}}$ and
$K_T = K_0 t^{-\tilde{\gamma}}$,
where $t=\vert (T - T_c) / T_c \vert$ is a reduced temperature.
The values of these critical exponents are known to be
$\tilde{\alpha} \sim  0.11$ and 
$\tilde{\gamma} \sim  1.2$, respectively.

At the critical point, the correlation length $\xi$ diverges
as $\xi =\xi_0t^{-\nu}$ with $\nu \sim 0.63$.
The bulk viscosity may also show a singular behavior around the QCD CP,
as shown in \cite{onuki:1997,Moore};
$\zeta = \zeta_0 t^{-a_{\zeta}}$
with $\zeta_0$ being a constant.
The exponent $a_{\zeta}$ for the liquid-gas transition
is predicted \cite{onuki:1997} to be $z\nu-\tilde{\alpha}$ with
 the dynamical critical exponent $z$;
 $z \sim 3$. Thus we have
$a_{\zeta}\sim 1.8$.
The singular behavior of the thermal conductivity around the critical point is
given by 
$a_{\kappa}$;
$\kappa =\kappa_0 t^{-a_{\kappa}}$
with $a_{\kappa}\sim 0.63$ where $\kappa_0$ is a constant.

Now using these formulae, we can show\cite{Minami:2009hn}
 that the width of the Rayleigh peak behaves as
$\Gamma_{\rm R}  \sim t^{\tilde{\gamma}-a_{\kappa}}$,
which tells us
 that the width $\Gamma_{\rm R}$ becomes narrow as the QCD CP is approached.   
We emphasize that this result is independent of the choice of the 
relativistic fluid dynamic equation or frame.

We find that the width  of the Brillouin peaks $\Gamma_{\rm B}$
show  the following critical behavior for the Landau and I-S equations, 
$\Gamma_{\rm B} \sim \frac{\zeta_0}{2 w_0} t^{-a_{\zeta}}$.
We note that this singularity comes from that of the bulk viscosity.    
In the case of TKO equation, we first note that 
the critical behavior of the effective bulk viscosity is given as
$\zeta^{'} = \frac{\zeta}{(3\gamma -4)^2} \sim t^{2\tilde{\gamma}-a_{\zeta}}$,
which shows that the effective bulk viscosity has a positive exponent and 
does not show a singular behavior
because $a_{\zeta} \sim 1.8$ and $\tilde{\gamma} \sim  1.2$.
Instead, the singularity of the Brillouin peaks for TKO equation 
comes from that of the thermal conductivity\cite{Minami:2009hn}; 
$\Gamma_{\rm B} \sim  t^{-(a_{\kappa}-\tilde{\alpha})}$.

Anyway, we have found that the width $\Gamma_{\rm B}$  diverges at the QCD CP, 
irrespective of the relativistic fluid dynamic equations,
although the strength of the singularity may differ depending on
the choice of the fluid dynamic equation\cite{Minami:2009hn}.

\begin{figure}[htbp]
\begin{tabular}{cc}
\begin{minipage}{0.5\hsize}
  \begin{center}
   \includegraphics[width=50mm,angle=-90]{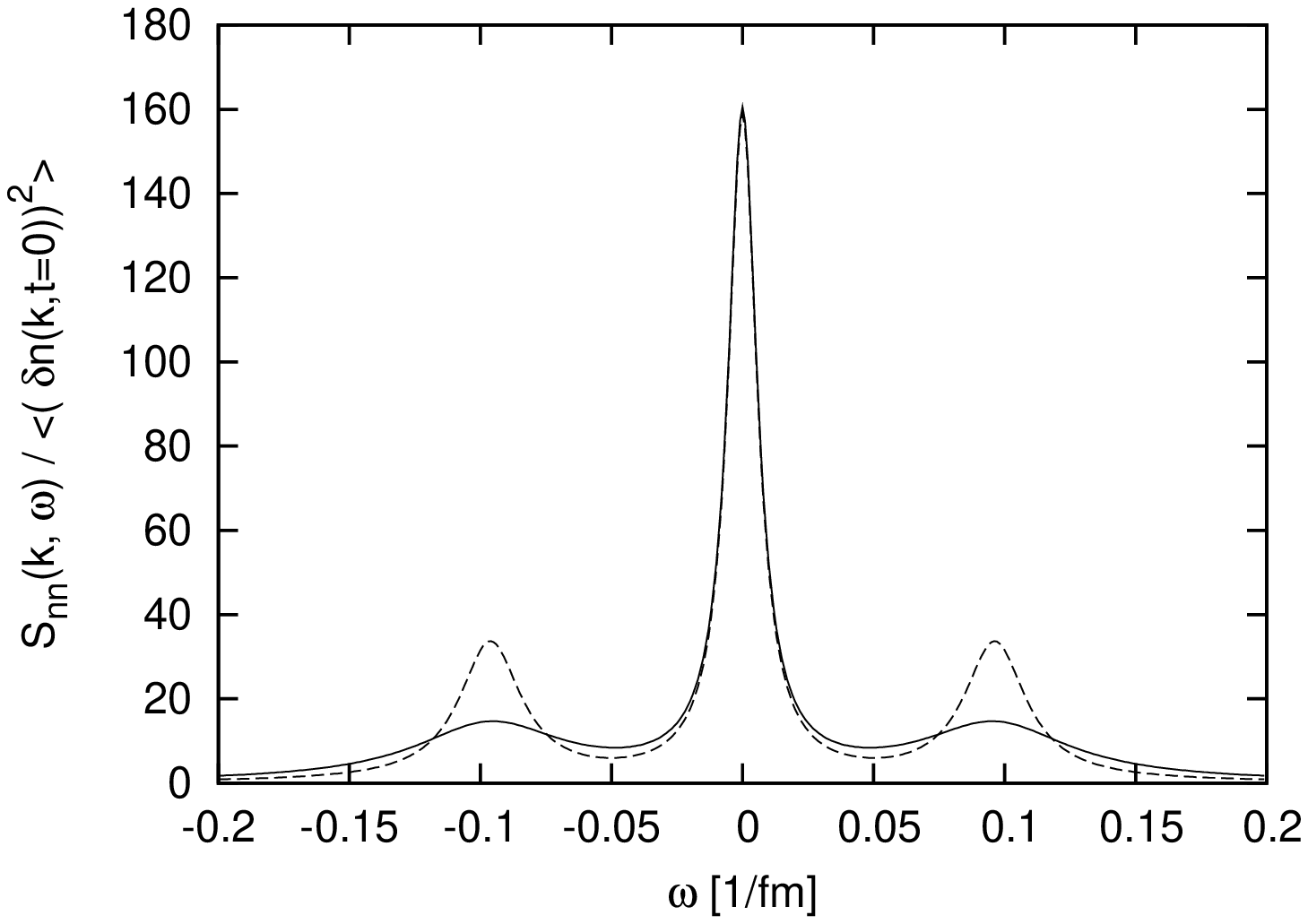}
\end{center}
\end{minipage}
\begin{minipage}{0.5\hsize}
 \begin{center}
   \includegraphics[width=50mm,angle=-90]{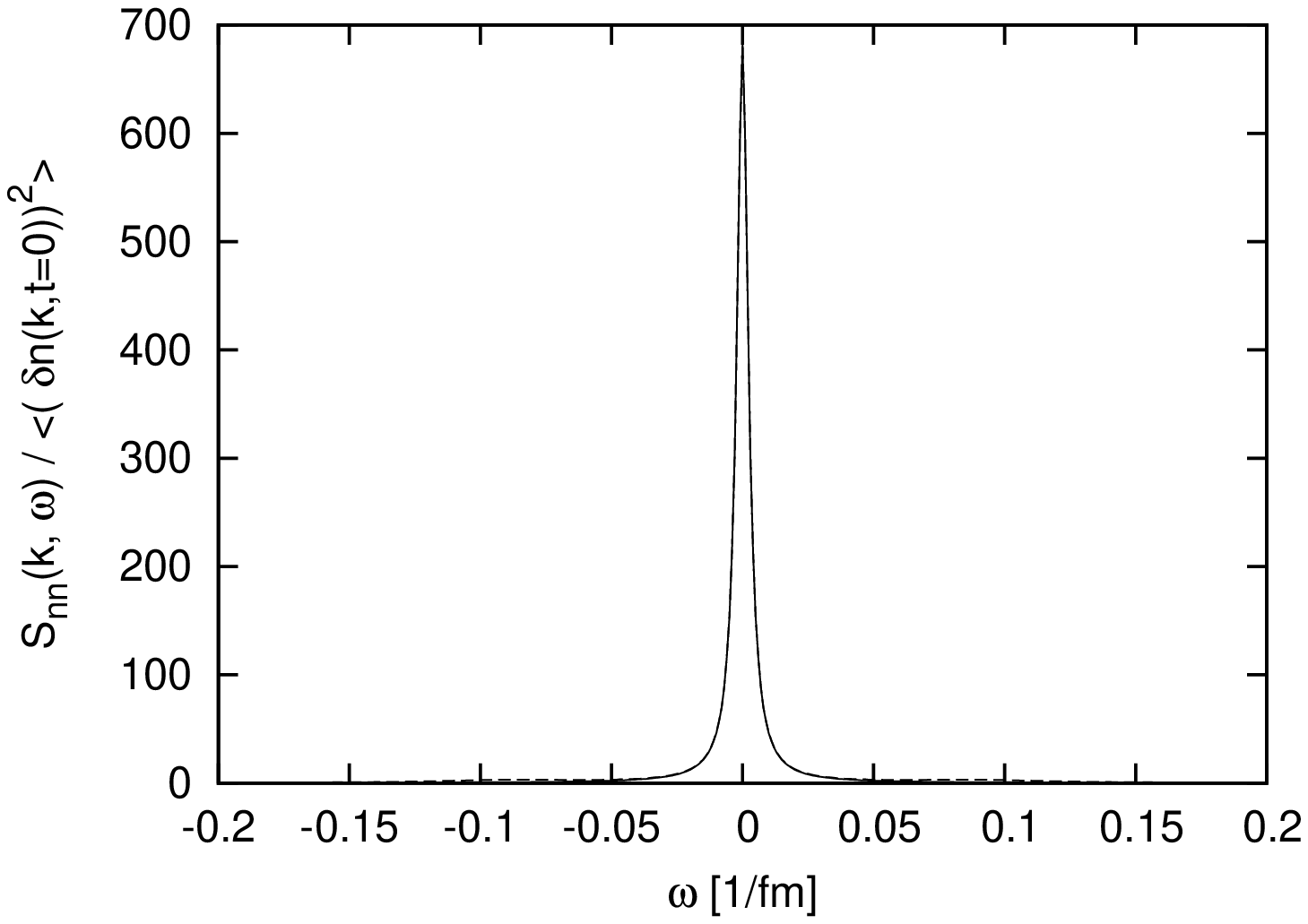}
\end{center}
\end{minipage}
\end{tabular}
  \caption{Left panel:\,
The spectral function at $t\equiv \vert T-T_c\vert/T_C=0.4$ and $k=0.1$\, [1/fm].
               The solid line represents the Landau and I-S cases.
           The dashed line represents the TKO case. 
           The strength of the Brillouin peaks becomes small due to the singularity of
 the ratio of specific heats.
 Right panel:The spectral function at $t=0.1$ and $k=0.1$\, [1/fm].
 We see that the Brillouin peaks which correspond to sound wave dies out
           and the difference between the Landau and TKO cases disappears. }
  \label{fig:t5-1}
%\end{figure}
%
%\begin{figure}[htb]
%  \includegraphics[width=45mm, angle=270]{t01.eps}
%   \includegraphics[width=45mm]{t01.eps}
%  \label{fig:t1}
%\end{figure}
\end{figure}

It turns out that
these singular behaviors of the width of the Brillouin peaks  around the
 QCD CP may not be observed:
Note that the strengths of the Rayleigh and the Brillouin peaks
are given in terms of $\gamma$, which
 behaves like 
$\gamma = \tilde{c}_p / \tilde{c}_n \sim t^{-\tilde{\gamma}+\tilde{\alpha}}
\rightarrow \infty$,
in the critical region. 
Then the strength of the Brillouin peaks is attenuated 
and only the Rayleigh peak stands out in the critical region, as follows;
\begin{eqnarray}
S_{n n}(\bfk ,\omega ) 
 \sim \la (\delta n(\bfk ,\, 0))^2\ra \frac{2\Gamma_{\rm R}
 k^{2}}{\omega^{2}+\Gamma_{\rm R}^{2}k^{4}}, \quad \quad 
(T\, \sim \, T_c).
\end{eqnarray}

 Figure~\ref{fig:t5-1} shows 
 how the dynamical structure factor for the density fluctuations
in the Landau (I-S) and TKO cases behaves around the 
QCD critical point
for $k=0.1$\, [1/fm],\, $t=0.5$ and $t=0.1$, respectively.
We see that the strength of the Brillouin peaks becomes small and dies out 
as the system approach the QCD CP. 
In addition, the static  correlation function $\la (\delta n(\bfk ,t=0))^2\ra$ 
shows a singular behavior in the forward direction $k=0$.
This is known as the critical opalescence\cite{stanley,reichl}.
Then the strength of the Rayleigh peak will be
most  drastically enhanced in the forward angle in the critical region.

Why at all do sound modes or mechanical density fluctuations die out at the critical point? 
To answer this question, we first note that
the fluid dynamic regime is expressed as
$\xi << \lambda_s$, where $\lambda_s$ represents the wave length of the sound mode.
When this condition is satisfied,
the sound mode can be developed\cite{Hohenberg:1977ym}.
However, in the vicinity of the critical point, the correlation length
$\xi$ becomes very large and eventually goes to infinity, so
the above inequality can not be satisfied, and the sound mode
is hardly developed in the vicinity of the critical point\cite{Hohenberg:1977ym}.
This is the reason why the sound mode or density fluctuation
is hardly developed around the QCD critical point.

\section{Possible disappearance or suppression of Mach cone around the QCD CP}

The result in the last section suggests
 that  phenomena inherently related to the existence
of the mechanical density fluctuations may disappear around the critical point.
One of such phenomena 
is the possible Mach cone formation \cite{Torrieri:2009mv} 
by the particle passing through 
the medium with a speed larger than the sound velocity $c_s$.
Such a Mach-cone like particle correlations 
are observed in the RHIC experiment\cite{:2008nd}. 
If such three-particle correlations have been 
confirmed to be a Mach-cone formation,
then the disappearance or suppression of
the Mach cone would be a signal that the created matter 
has passed through the critical region, showing the existence of
the QCD critical point.
Even if the thermal wake also contributes
to the formation of Mach cone, a suppression of Mach cone may be 
expected by the attenuation of the dynamical density fluctuations. 
So it would be very interesting to see possible
variation of the strength of the Mach cone 
according to the variation of the incident energy of the heavy-ion collisions.
In theoretical side, it is an intriguing task to explore 
the fate of Mach cone with an equation of state 
that admits the existence of the CP.

\section{Summary and concluding remarks}

Motivated by the fact that QCD critical point (CP) belongs to
the same universality class as the liquid-gas transition,
we have analyzed the dynamical density fluctuations using relativistic
fluid dynamics.
%It was shown that  the relativistic effects on the spectral function, or
%the dynamical structure factor, 
%of the density fluctuation appears only in the width of the Lorentzian peaks
%corresponding to the sound and thermal modes
%through the modification of the transport/kinetic coefficients.
We have found that 
the mechanical density fluctuations are attenuated owing to the divergence of
the correlation length $\xi$ around the QCD CP;
on the other hand, the entropy fluctuation in turn gets enhanced
and tends to makes a single peak around the QCD CP in the dynamical structure
factor of the density fluctuations.
Sunch an attenuation of the mechanical density fluctuations
 may lead to a suppression or even total disappearance
 of  Mach cone at the QCD CP.
 If the Mach cone formation is confirmed at the incident
energy $\sqrt{s_{NN}}= 200$ GeV in RHIC experiments,
 possible disappearance or strong suppression of a Mach cone along 
with the lowering
of the incident energy can be a signal of the existence of the critical point;
the created matter should have gone through the critical region of the CP.

Eexplicit calculations with equation of motion which admittes the existence
of the critical point is necessary for confirm the fate of Mach cone formation. 
To make a direct connection with RHIC experiment, we should analyze the
density fluctuations with the expanding back ground\cite{progress}.

As a future work, we should study the coupling between 
the thermal fluctuations and the transverse mode
using the mode-mode coupling theory
in the close vicinity of the critical point for the relativistic case.

Finally, we would like to indicate that
there are still other possibilities in the structure of  the QCD phase diagram
\cite{Zhang:2008wx,Hatsuda:2006ps,Fukushima:2008is}: For example, there may
exist multiple critical points when the color superconductivity and 
the vector interaction like $g_{_V}(\bar{q}\gamma^{\mu}q)^2$ 
\cite{Zhang:2008wx} and/or the anomaly term\cite{Hatsuda:2006ps} are incorporated.
It suggests that the QCD matter is very soft along the critical line 
when the color superconductivity is incorporated, which may imply that
there are large fluctuations of various observables including
  diquark-density mixed ones.
% An analysis of such a possibility
%may involve a {\em dynamical} Hartre(-Fock)-Bogoliubov theory in the
%relativistic kinematics, which should be an interesting theoretical challenge
%worthwhile to explore.

T.K. is grateful to Krishna Rajagopal for his valuable comments. 
This work was partially supported by a
Grant-in-Aid for Scientific Research by the Ministry of Education,
Culture, Sports, Science and Technology (MEXT) of Japan (Nos.
20540265, 19$\cdot$07797),
 by Yukawa International Program for Quark-Hadron Sciences, and by the
Grant-in-Aid for the global COE program `` The Next Generation of
Physics, Spun from Universality and Emergence '' from MEXT.


\begin{thebibliography}{99}

\bibitem{qcdcp}
%\bibitem{Asakawa:1989bq}
  M.~Asakawa and K.~Yazaki,
{\it Chiral restoration at finite density and temperature},
  Nucl.\ Phys.\  A {\bf 504} (1989) 668;\,
%\bibitem{Barducci:1989}
 A.~Barducci, R.~Casalbuoni, S.~De Curtis, R.~Gatto and G.~Pettini,
  {\it Chiral symmetry breaking in QCD at finite temperature and density},
  Phys.\ Lett.\  B {\bf 231} (1989) 463;  \,
 Phys.\ Rev.\  D {\bf 41} (1994) 1610.
%
\bibitem{Ejiri:2007ga}
  S.~Ejiri,
  {\it On the existence of the critical point in finite density lattice QCD},
  Phys.\ Rev.\  D {\bf 77} (2008), 014508;\,
%S.~Ejiri,
  {\it Canonical partition function and finite density phase transition in lattice
  QCD},
  Phys.\ Rev.\  D {\bf 78} (2008) 074507.
%
\bibitem{Zhang:2008wx}
  M.~Kitazawa, T.~Koide, T.~Kunihiro and Y.~Nemoto,
  {\it Chiral and color superconducting phase transitions with vector interaction
  in a simple model},
  Prog.\ Theor.\ Phys.\  {\bf 108} (2002) 929;\, 
  Z.~Zhang, K.~Fukushima and T.~Kunihiro,
  {\it Number of the QCD critical points with neutral color superconductivity},
  Phys.\ Rev.\  D {\bf 79} (2009) 014004;\,
%  [arXiv:0808.3371 [hep-ph]].
  %%CITATION = PHRVA,D79,014004;%%
Z.~Zhang and T.~Kunihiro,
{\it Vector interaction, charge neutrality and multiple chiral critical point
  structures},
 Phys. Rev. D {\bf 80} (2009) 014015.
%
\bibitem{Hatsuda:2006ps}
  T.~Hatsuda, M.~Tachibana, N.~Yamamoto and G.~Baym,
{\it New critical point induced by the axial anomaly in dense QCD},
  Phys.\ Rev.\ Lett.\  {\bf 97} (2006) 122001.
%  [arXiv:hep-ph/0605018];
%\bibitem{Yamamoto:2007ah}
%  N.~Yamamoto, M.~Tachibana, T.~Hatsuda and G.~Baym,
%{\it Phase structure, collective modes, and the axial anomaly in dense QCD,''
%  Phys.\ Rev.\  D {\bf 76} (2007), 074001.
%
\bibitem{Fukushima:2008is}
 P.~de Forcrand and O.~Philipsen,
  {\it The chiral critical line of $N_f=2+1$ QCD at zero and non-zero baryon
  density},  JHEP {\bf 0701} (2007) 077;\,
% P.~de Forcrand and O.~Philipsen,
  %``The chiral critical point of Nf=3 QCD at finite density to the order
  %(mu/T)^4,''
%  JHEP {\bf 0811} (2008), 012.
%O.~Philipsen,
  %``Status of Lattice Studies of the QCD Phase Diagram,''
%  Prog.\ Theor.\ Phys.\ Suppl.\  {\bf 174} (2008), 206;\\
 K.~Fukushima,
{\it Critical surface in hot and dense QCD with the vector interaction},
  Phys.\ Rev.\  D {\bf 78} (2008) 114019.
%
\bibitem{Hohenberg:1977ym}
  P.~C.~Hohenberg and B.~I.~Halperin,
{\it Theory of Dynamic Critical Phenomena},
  Rev.\ Mod.\ Phys.\  {\bf 49} (1977) 435.
%
\bibitem{fujii}
  H.Fujii,{\it Scalar density fluctuation at critical end point in NJL model},
Phys.\ Rev\ .D {\bf 67} (2003) 094018;\,
  H.Fujii and M.Ohtani,
{\it Sigma and hydrodynamic modes along the critical line},
 Phys.\ Rev.\ D {\bf 70} (2004) 014016.
% H.~Fujii and M.~Ohtani,
  %``Soft modes at the critical end point in the chiral effective models,''
%  Prog.\ Theor.\ Phys.\ Suppl.\  {\bf 153} (2004), 157;\\
%H.~Fujii and N.~Tanji,
  %``Soft mode of the QCD critical point,''
%  J.\ Phys.\ G {\bf 35} (2008), 104060.
%
\bibitem{son}
  D.T.Son~and~M.A.Stephanov,
{\it Dynamic universality class of the QCD critical point},
Phys.\ Rev.\ D {\bf 70} (2004) 056001.
%
\bibitem{Kunihiro:1991qu}
  T.~Kunihiro,
{\it Quark number susceptibility and fluctuations in the vector channel at high
  temperatures},
  Phys.\ Lett.\  B {\bf 271} (1991) 395.
%
\bibitem{Kunihiro:2000ap}
  T.~Kunihiro,
{\it Chiral transition and baryon-number susceptibility},
{\em CONFINEMENT 2000}, Edited by H. Suganuma, M. Fukushima, H. Toki,
( Singapore, World Scientific, 2001) [{\tt arXiv:hep-ph/0007173}].
%
\bibitem{Hatta:2002sj}
  Y.~Hatta and T.~Ikeda,
{\it Universality, the QCD critical / tricritical point and the quark number
  susceptibility},
  Phys.\ Rev.\  D {\bf 67} (2003) 014028.
%
\bibitem{Moore}
 G.~D.~Moore and O.~Saremi,
{\it Bulk viscosity and spectral functions in QCD},
  JHEP {\bf 0809} (2008) 015.
%
\bibitem{onuki:1997}
A.~Onuki, 
{\it Dynamic equations and bulk viscosity near the gas-liquid critical point}, 
Phys.~Rev.~E \textbf{55} (1997) 403.
%
\bibitem{Minami:2009hn}
  Y.~Minami and T.~Kunihiro,
{\it Dynamical Density Fluctuations around QCD Critical Point Based on
  Dissipative Relativistic Fluid Dynamics --- possible fate of Mach cone at the
  critical point---},
 Prog. Theor. Phys. {\bf 122} (2009) 881; [{\tt arXiv:0904.2270[hep-th]}]. 
%
\bibitem{stanley}
H.~E.~Stanley, {\it Introduction~to~Phase~Transitions~and~Critical
Phenomena},(Oxford University Press,~Oxford,~1977).
%
\bibitem{:2008nd}
As a latest reference, see  B.~I.~Abelev {\it et al.}~[STAR Collaboration],
  {\it Indications of Conical Emission of Charged Hadrons at RHIC},
  Phys.\ Rev.\ Lett.\  {\bf 102} (2009) 052302.
%
\bibitem{Torrieri:2009mv}
  H.~Stoecker,
{\it Collective Flow signals the Quark Gluon Plasma},
  Nucl.\ Phys.\  A {\bf 750} (2005) 121;\,
L.~M.~Satarov, H.~Stoecker and I.~N.~Mishustin,
{\it Mach shocks induced by partonic jets in expanding quark-gluon plasma},
  Phys.\ Lett.\  B {\bf 627} (2005) 64;\,
 J.~Casalderrey-Solana, E.~V.~Shuryak and D.~Teaney,
{\it Conical flow induced by quenched QCD jets},
  J.\ Phys.\ Conf.\ Ser.\  {\bf 27} (2005) 22
  [Nucl.\ Phys.\  A {\bf 774} (2006) 577].
%
\bibitem{landau}
  L.~D.~Landau~and~E.~M.~Lifshitz,~{\it Fluid~Mechanics}~(Pergamon,New~York,~1959).
%
\bibitem{eckart}
  C.~Eckart, {\it The Thermodynamics of irreversible processes. 3. Relativistic theory of the
  simple fluid},
Phys.\ Rev.\ {\bf 58} (1940) 919.

\bibitem{hiscock}
  W.~A.~Hiscock and L.~Lindblom,
{\it Generic instabilities in first-order dissipative relativistic fluid
  theories},
 Phys.\ Rev.\ D {\bf 31} (1985) 725.  
%
\bibitem{is}
  W.~Israel,
{\it Nonstationary Irreversible Thermodynamics: A Causal Relativistic Theory },
Ann.Phys.(N.Y.) {\bf 100} (1976) 310;\,
  W.~Israel and ~J.M.Stewart,
{\it Transient relativistic thermodynamics and kinetic theory},
Ann. Phys.(N.Y.) {\bf 118}(1979) 341.
%
\bibitem{tko}
  K.~Tsumura, T.~Kunihiro and K.~Ohnishi,
{\it Derivation of covariant dissipative fluid dynamics in the
  renormalization-group method},
  Phys.\ Lett.\  B {\bf 646} (2007) 134.
%
\bibitem{tk}
  K.~Tsumura,~T.~Kunihiro,
{\it Stable First-order Particle-frame Relativistic Hydrodynamics for
  Dissipative Systems },
Phys.\ Lett.\ B {\bf 668} (2008) 425.
%  [arXiv:0709.3645 [nucl-th]].
\bibitem{mountain}
 L.~P.~Kadanoff and P.~C.~Martin,
{\it Hydrodynamic Equations and Correlation Functions},
 Ann.\ Phys.\ {\bf 24} (1963) 419;\,
  R.~D.~Mountain,
{\it  Spectral Distribution of Scattered Light in a Simple Fluid},
 Rev.\ Mod.\ Phys.\ {\bf 38}(1966) 205.
%
\bibitem{reichl}
%As a comprehensive account of the light scattering by
%the fluid dynamical density fluctuations see,
  L.~E.~Reichl,~{\it A Modern Course in Statistical Physics}(Wiley-Interscience 1998).
%
\bibitem{progress} K.~ Tsumura, Y.~ Minami and T.~Kunihiro, in progress.
\end{thebibliography}
\end{document}